\documentclass[aip,apl,amssymb,amsmath,reprint]{revtex4-1}
\usepackage{bm}
\usepackage{graphicx}
\usepackage{braket}
\usepackage{hyperref}
\usepackage[all]{hypcap}
\hypersetup{
  colorlinks   = true, 
  urlcolor     = blue, 
  linkcolor    = red, 
  citecolor   = blue 
}
\begin{document}

\title{A 2$\times$2 quantum dot array with controllable inter-dot tunnel couplings}

\author{Uditendu Mukhopadhyay}
\thanks{These authors contributed equally to this work}
\affiliation{QuTech and Kavli Institute of Nanoscience, TU Delft, 2600 GA Delft, The Netherlands}

\author{Juan Pablo Dehollain}
\thanks{These authors contributed equally to this work}
\affiliation{QuTech and Kavli Institute of Nanoscience, TU Delft, 2600 GA Delft, The Netherlands}

\author{Christian Reichl}
\affiliation{Solid State Physics Laboratory, ETH Z\"{u}rich, Z\"{u}rich 8093, Switzerland}

\author{Werner Wegscheider}
\affiliation{Solid State Physics Laboratory, ETH Z\"{u}rich, Z\"{u}rich 8093, Switzerland}

\author{Lieven M. K. Vandersypen}
\email{L.M.K.Vandersypen@tudelft.nl}
\affiliation{QuTech and Kavli Institute of Nanoscience, TU Delft, 2600 GA Delft, The Netherlands}

\date{\today}

\begin{abstract}
The interaction between electrons in arrays of electrostatically defined quantum dots is naturally described by a Fermi-Hubbard Hamiltonian. Moreover, the high degree of tunability of these systems make them a powerful platform to simulate different regimes of the Hubbard model. However, most quantum dot array implementations have been limited to one-dimensional linear arrays. In this letter, we present a square lattice unit cell of 2$\times$2 quantum dots defined electrostatically in a AlGaAs/GaAs heterostructure using a double-layer gate technique. We probe the properties of the array using nearby quantum dots operated as charge sensors. We show that we can deterministically and dynamically control the charge occupation in each quantum dot in the single- to few-electron regime. Additionally, we achieve simultaneous individual control of the nearest-neighbor tunnel couplings over a range 0-40~$\mu$eV. Finally, we demonstrate fast ($\sim 1$~$\mu$s) single-shot readout of the spin state of electrons in the dots, through spin-to-charge conversion via Pauli spin blockade. These advances pave the way to analog quantum simulations in two dimensions, not previously accessible in quantum dot systems.
\end{abstract}

\keywords{Quantum dots, Tunnel coupling, Hubbard model}

\maketitle

Electrostatically defined quantum dots in semiconductors have been proposed as the basic underlying hardware in quantum computation~\cite{LD}, as well as digital and analog quantum simulations~\cite{Yamamoto,DasSarma,Stafford,Piere}. This is due to their ease of tunability, control of the relevant  parameters, fast measurement of the spin and charge states, and their potential for scalability. In particular, quantum dot arrays are natural candidates for simulating the Fermi-Hubbard model, as they adhere to the same Hamiltonian:
\begin{eqnarray}
H &=& \sum_i U_i n_{i\uparrow} n_{i\downarrow} - \sum_{i,j,\sigma} t_{i,j} \left(c^\dagger_{i\sigma} c_{j\sigma} + h.c.\right) - \sum_i \mu_{i} n_{i}\nonumber\\
& &+ \sum_{i,j} V_{i,j} n_i n_j
\label{eq:1}
\end{eqnarray}
The on-site interaction energy $U_i$ corresponds to the quantum dot charging energy on site $i$ and the hopping energy $t_{i,j}$ corresponds to the tunnel coupling between dots $i$ and $j$. The chemical potential term $\mu_i$ controls the electron number in each dot, as well as the relative energy detuning between dots. For quantum dot arrays there is an additional term $V_{i,j}$ that describes the inter-site Coulomb interaction energy. The operators $c_i, c_i^{\dagger}, n_i$ in Eqs.~\ref{eq:1} represent the second quantization annihilation, creation and number operators respectively, with the individual spins of the electrons are denoted by the subscript $\sigma = \{ \uparrow,\downarrow \}$. For simplicity, we have assumed that no external magnetic field is present in the system.

For the study of Fermi-Hubbard physics, control of the ratio $U/t$ is essential~\cite{Imada,Lee,Balents}. The hopping term can be tuned electrostatically, covering a range $t \approx 0-100$~$\mu$eV between nearest neighbors in a linear array~\cite{Hensgens}. The on-site interaction energy $U$ is set by the shape of the confinement potential and is not freely tunable, but it can be accurately measured with typical values of 1~-~10~meV~\cite{Kouwenhoven1997}. Similarly, $V$ is not tunable independently but can be measured precisely.

Quantum simulations of the Fermi-Hubbard model have previously been explored experimentally in cold atom systems~\cite{J2008,Parsons,Boll,Cheuk,Mazurenko}, manipulating arrays of the order of 100 atoms. However, these experiments are often limited by the initial entropy of the system~\cite{Parsons,Boll,Cheuk}. Quantum dot arrays can overcome this problem by operating in dilution refrigerators, where electron temperatures can reach $kT_e \sim 1$~$\mu$eV. On the other hand, experiments with quantum dots are still mainly being performed with linear arrays with no more than a few sites~\cite{Hensgens,Flentje2017,Malinowski2018}. Efforts to go beyond 1D with quantum dot arrays have so far stopped short of achieving well-characterized tunnel couplings in the few-electron regime~\cite{Tri,Tristan,Tarucha}.

In this letter we report on the design, fabrication and measurement of a quantum-dot plaquette in a 2$\times$2 geometry. We describe a fabrication technique used to implement a two-layer gate structure needed for this device. We then present measurements that demonstrate deterministic filling of electrons in all dots and controllable tunnel coupling over a large range (0~-~40~$\mu$eV) between all nearest-neighbor pairs. As the final ingredient for this quantum simulator, we perform single-shot measurements of the two-electron singlet/triplet states ($\ket{S}$/$\ket{T}$) using two dots in the array.

\begin{figure}
\begin{center}
 \includegraphics{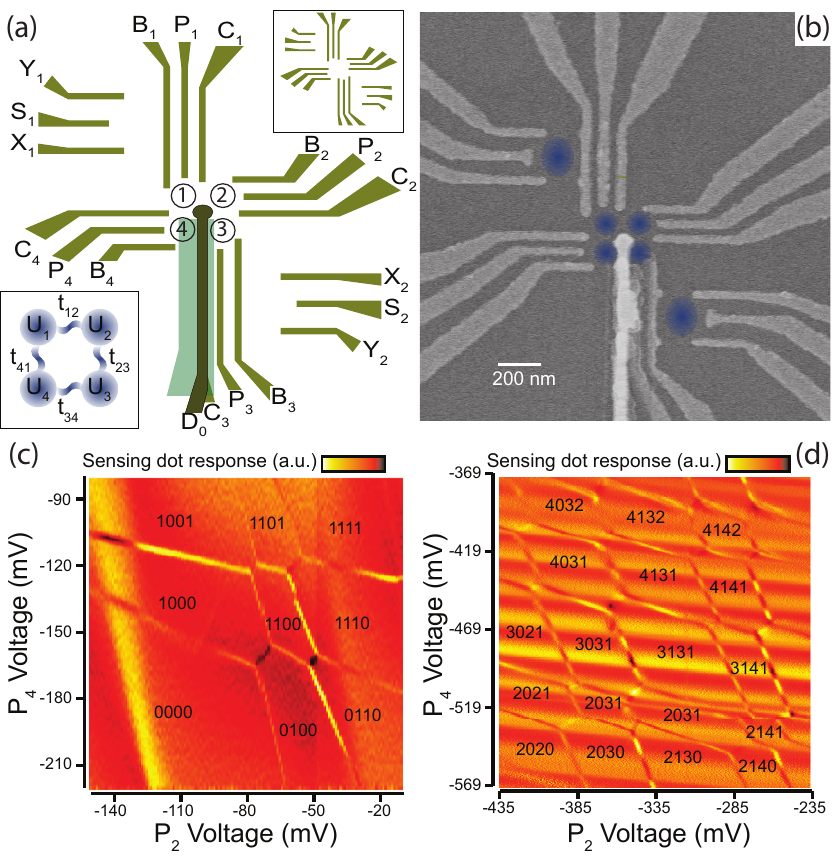}
\caption{(a) Schematic of the gate design, with the dot locations labeled in the center. (first layer in the top inset, bottom inset shows a schematic of the dot plaquette, with relevant Hubbard model terms). (b) SEM image of a device from the same batch as the one used for measurements. The overlaid blue circles are impressions of the dot wave-functions. (c)-(d) Charge stability diagrams showing controlled filling of all four quantum dots in the single- (c) and few- (d) electron regime. The data in (c) and Fig.~\ref{fig:3} was taken in one device cooldown, the data in panel (d), Table I and Fig.~\ref{fig:2} in another cooldown.} \label{fig:1}
\end{center}
\end{figure}

The device contains electrostatically defined quantum dots formed by selectively depleting electrons using nano-fabricated gate electrodes on the surface of a GaAs/AlGaAs heterostructure. The gate pattern is designed to form four quantum dots in a 2$\times$2 geometry, where the nearest neighbors are cyclic, i.e. $i+4=i$ [Fig.~\ref{fig:1}(a)]. The coupling of each of the dots to its own electron reservoir is controlled through the constriction created between the B$_{i+1}$ and the C$_i$ gates. This is designed to allow for operation of the quantum dots in the isolated regime~\cite{Flentje2017,Bertrand2015}; however, we do not explore this configuration here. Deterministic electron filling of the quantum dots is achieved by adjusting $\mu_i$ relative to the Fermi energy of the reservoirs, through the use of the gates P$_i$. A center gate (D$_0$) reaches the substrate at the center of the plaquette. Biasing this gate negatively effectively separates the dots from each other. It thereby suppresses tunnel couplings along the two diagonals of the array and also influences the nearest-neighbor tunnel couplings (along the perimeter of the array), since the combination of D$_0$ with a C$_i$ gate controls $t_{i,i+1}$. The device design also includes an extra set of gates (X$_i$, Y$_i$, S$_i$) used to define two larger dots to be operated as charge sensors. The GaAs/Al$_x$Ga$_{1-x}$As heterostructure is Si-doped, with a two-dimensional electron gas at the 90 nm deep interface (x = 0.314, mobility = 1.6x10$^6$cm$^{2}$/Vs  and electron density = 1.9x10$^{11}$cm$^{-2}$).  All gates except D$_0$ are fabricated in a first layer of Ti/Au of thickness 5/20 nm, evaporated on the bare substrate and patterned following standard procedures\cite{ThesisTim} (the top inset in Fig.~\ref{fig:1}(a) shows the schematic of this layer). The D$_0$ gate runs above gate C$_3$ and contacts the substrate at the center of the array with a foot of $\sim 50$ nm radius. It is fabricated using 10/100 nm evaporated Ti/Au and isolated from the bottom layer gates using a 50 nm thick, 200 nm wide and 1.5~$\mu$m long dielectric slab of SiN$_x$, fabricated using sputtering and lift-off. For this step, an 80 nm thick layer of AR-P 6200\cite{CSAR} is used as the e-beam resist and lift-off is performed in hot (80$^o$C) N-Methyl-2-Pyrrolidone. A scanning electron microscope (SEM) image of a completed device is shown in Fig.~\ref{fig:1}(b).

\begin{table*}[t]
\caption{Relevant gate voltages and lever arms}\label{table:I}
\bgroup
\def\arraystretch{1.5}
\resizebox{\textwidth}{!}{%
\begin{tabular}{c c c c c c c c c c c c c c c c c c c c c}
\hline
\hline
  & B1 & B2 & B3 & B4 & P1 & P2 & P3 & P4 & C1 & C2 & C3 & C4 & D0 & X1 & X2 & Y1 & Y2 & S1 & S2\\
 \hline
 \hline
 Voltages at 1111 (mV) & -150 & -230 & -130 & -100 & -263 & -60 & -9 & -221 & -120 & -180 & -180 & -220 & -180 & -360 & -120 & -280 & -270 & -110 & -390 \\
 \hline
 Voltages at 3131 (mV)  & -100 & -20 & -90 & -194 & -169 & -335 & -30 & -469 & -188 & -141 & -37 & -57 & -135 & -343 & -95 & -310 & -274 & -429 & -504 \\
 \hline
 Bias cooling voltage (mV) & 300 & 250 & 300 & 250 & 150 & 150 & 150 & 150 & 250 & 250 & 250 & 250 & 200 & 350 & 350 & 300 & 300 & 200 & 200   \\
 \hline
 Lever Arms ($\mu$eV/mV)  &  &  & & & 39 & 41 & 54 & 31  \\
 (Plungers to dots)   &  &  & & & (D1) & (D2) &  (D3) & (D4) \\
 \hline
 \hline
\end{tabular}}
\egroup
\end{table*}

The device was cooled down with positive bias voltages (see values in Table~\ref{table:I}) on all gates in order to decrease charge noise~\cite{BiasCooling}. All the P$_i$ and C$_i$ gates are connected to high-frequency ($\sim 1$~GHz) lines for pulsing and fast sweeping. One reservoir for each sensing dot is connected to a resonant RF circuit for high-bandwidth (up to 3~MHz) charge sensing. The two readout circuits have resonance frequencies of 84.5 and 130.6~MHz, are connected to a single amplifier chain and are read out simultaneously using frequency multiplexing~\cite{RF}. By measuring charge stability diagrams using different combinations of gates, we can identify and tune the four dots to the few-electron regime. In Figs.~\ref{fig:1}(c,d), we show examples of two charge stability diagrams, where we have identified the charge states of the four dots, ranging from (0000) to (4142), where $(klmn)$ indicate the charge occupation of dots 1 through 4. The different cross-capacitances between the dots and the gates lead to charge transition lines with four different slopes in the charge stability diagrams, corresponding to the filling of the four dots.

Using these diagrams, appropriate voltages can be applied to the gates to achieve deterministic filling of the dots. Although we can reach the regime with one electron in each dot, it was difficult to tunnel couple all neighboring dots. We attribute this to the center gate being slightly too large. To bypass this problem, we keep the first orbital shells of dots 1 and 3 filled with two electrons each. In this configuration the electron wavefunction is larger, which facilitates tunnel coupling neighboring dots. However, it is important to note that in this configuration, the unpaired electron occupies an antisymmetric (2p) orbital~\cite{atom}, which can result in effects such as a sign inversion in the tunnel coupling. The gate voltages needed to achieve (1111) and (3131) charge states are specified in Table~\ref{table:I}. We perform finite voltage-bias measurements~\cite{VanderWiel2002,House} to extract the lever arm (see Table~\ref{table:I}) between gate voltage and dot chemical potential energy. Using these, the charging energies for the four dots are then estimated from the distance between charge transition lines in the charge stability diagrams [$U_1 = 2.1$~meV, $U_3 = 2.3$~meV (3 electron dots) and $U_2 = 3.4$~meV, $U_4 = 3.3$~meV (1 electron dots)]. From the same diagrams we also extract the inter-site Coulomb interaction energies $V_{1,2} = 0.67$~meV, $V_{2,3} = 0.55$~meV, $V_{3,4} = 0.47$~meV, $V_{4,1} = 0.39$~meV.

\begin{figure}
\begin{center}
\includegraphics{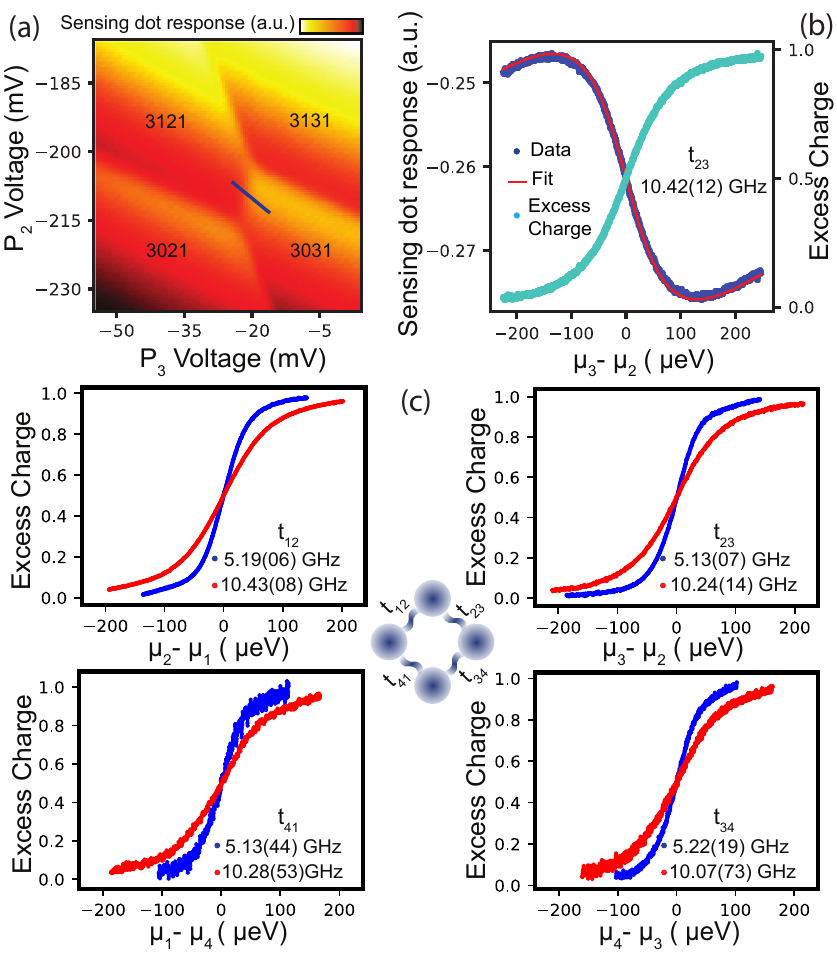}
\caption{(a) Charge stability diagram zoomed in on an inter-dot transition. (b) A line cut of panel (a) along the detuning axis (blue line in (a)) and fitting of the line to get tunnel coupling and excess charge distribution. (c) Excess charge extracted from the sensing dot signal when changing gate voltages along the detuning axis for the four different double dots in the plaquette. The data shows controllable tunnel couplings between all nearest-neighbor double-dot pairs. All the curves of the same color were taken using the same global gate configuration.} \label{fig:2}
\end{center}
\end{figure}

We next characterize and control the four inter-dot tunnel couplings. Starting with the array in the (3131) charge state, we measure $t_{i,j}$ by moving to a gate voltage configuration that removes one electron from the system and is centered at $\mu_{i} = \mu_{j}$ while keeping the other two dots (slightly) detuned. Around this point, the charge stability diagram shows an inter-dot transition line [Fig.~\ref{fig:2}(a)]. As we sweep the voltage along the detuning axis (perpendicular to the inter-dot transition), the charge sensor signal displays a step as the extra electron moves over from one dot to the other. The width of this step is dependent on the tunnel coupling $t_{i,j}$ and the electron temperature $T_e$~\cite{DiCarlo,Hensgens}. Fig.~\ref{fig:2}(b) shows a sample measurement where the sensor signal is plotted as we sweep the gate voltages across the inter-dot transition. This signal is then fitted to extract $t_{i,j}$ given $T_e \sim 70$~mK ($\sim 6$~$\mu$eV). $T_e$ was measured by fitting a similar trace for the case $t << T_e$. Note that this measurement of $T_e$ provides an upper bound for the charge noise. From the fits to the current traces, we derive the excess charge as a function of detuning between the two dots [Fig.~\ref{fig:2}(b)].

Nearest-neighbor tunnel couplings can be controlled electrostatically by opening/closing the constrictions created between D$_0$ and the C$_i$ gates. However, if we vary these gates only, the cross-capacitance between these gates and the dots result in unwanted changes in the chemical potential of the dots. To remedy this, we map out a cross-capacitance matrix that expresses the capacitive coupling between all gates and every dot. For small changes in gate voltage ($< \sim 100$ mV) we can assume these cross-capacitances to remain constant and the changes in the individual dot energies can be expressed as linear combinations of gate voltages:
\begin{eqnarray}
\begin{bmatrix}
\delta\mu_1 & \delta\mu_2 & \delta\mu_3 & \delta\mu_4
\end{bmatrix}
=
\delta\bm{G \alpha} \\
\bm{G} = \begin{bmatrix}
P_1 & P_2 & P_3 & P_4 & C_1 & C_2 & C_3 & C_4 & D_0 \nonumber
\end{bmatrix}
\end{eqnarray}
where $\bm{\alpha}$ is a $4\times9$ matrix of cross-capacitances: $\alpha_{i,i}$ corresponds to the lever-arm of gate P$_i$ to dot $i$, and $\alpha_{i,j} = \alpha_{i,i} \beta_{i,j}$, where $\beta_{i,j} = \delta P_i/\delta G_j$ is the slope of the charge transition of dot $i$, which can be extracted from a charge stability diagram. Once extracted, $\bm{\alpha}$ can then be used to define virtual gates~\cite{Hensgens} (C$'_i$ or D$'_0$) that allow us to vary one of the C$_i$ or D$_0$ gates, while simultaneously adjusting all the P$_i$ gates to keep $\delta \mu_i = 0$. For example, for C$'_i$ the adjustment of P$_i$ can be calculated from:
\begin{equation}
\begin{bmatrix}
\delta P_1 \\ \delta P_2 \\ \delta P_3 \\ \delta P_4
\end{bmatrix}
= -\delta C_1
\begin{bmatrix}
\alpha_{1,1} & \alpha_{1,2} & \alpha_{1,3} & \alpha_{1,4} \\
\alpha_{2,1} & \alpha_{2,2} & \alpha_{2,3} & \alpha_{2,4} \\
\alpha_{3,1} & \alpha_{3,2} & \alpha_{3,3} & \alpha_{3,4} \\
\alpha_{4,1} & \alpha_{4,2} & \alpha_{4,3} & \alpha_{4,4}
\end{bmatrix}^{-1}
\begin{bmatrix}
\alpha_{1,5} \\ \alpha_{2,5} \\ \alpha_{3,5} \\ \alpha_{4,5}
\end{bmatrix}
\end{equation}

This technique significantly simplified the process of adjusting the tunnel barriers and was a key element in achieving effective tunnel coupling control. In Fig.~\ref{fig:2}(c) this control is demonstrated by uniformly setting all four tunnel couplings to 5~GHz ($\sim 20$~$\mu$eV, blue traces) and 10~GHz ($\sim 40$~$\mu$eV, red traces).

\begin{figure}
\begin{center}
\includegraphics{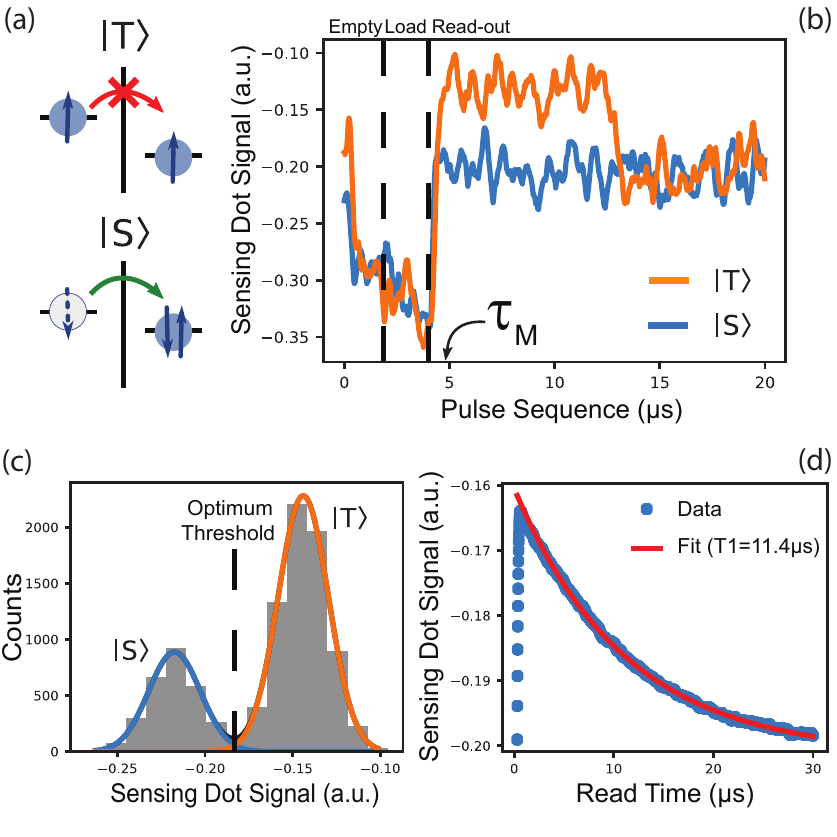}
\caption{(a) Schematic of the spin to charge conversion process used to read out the spin states via Pauli Spin Blockade. (b) Example single-shot read-out traces for singlet (blue) and triplet (orange) states. (c) A histogram of the current signal at time $\tau_\textsc{m}$ constructed from 10000 single-shot measurements. Solid lines are Gaussian fits to the two peaks in the histogram corresponding to singlet (blue) and triplet (orange) states. (d) Average signal obtained from 10000 read-out traces. Solid line is an exponential fit, from which we extract the relaxation time $T_1$.}\label{fig:3}
\end{center}
\end{figure}

Finally we demonstrate single-shot read-out of two-spin states using a three-stage pulse~\cite{Elzerman}. The Pauli exclusion principle~\cite{Pauli} is used to convert a charge measurement into a measurement that distinguishes between singlet and triplet states of two spins occupying neighboring quantum dots. We follow a protocol used previously to read out spins in a double dot~\cite{PSB} where a random two-spin state is loaded in the (1,1) charge configuration. The detuning between the dots is then pulsed to favor tunneling towards the (2,0) charge state. For a singlet ($\ket{S}$), tunneling to (2,0) is allowed. For a triplet ($\ket{T}$) however, the Pauli exclusion principle requires the (2,0) state to occupy the first excited state orbital of the dot, which is energetically inaccessible ($\sim 0.4$~meV away). Therefore, spins in $\ket{T}$ remain in the (1,1) state [Fig.~\ref{fig:3}(a)] until they relax to $\ket{S}$, with rate $1/T_1$. To identify the spin states, we monitor the charge sensor signal at a specific time $\tau_\textsc{m}$ after the start of the read-out pulse. We integrate the signal for 0.1~$\mu$s around $\tau_\textsc{m}$. If the integrated signal exceeds (does not exceed) a fixed threshold, we conclude the charge state was (1,1) [(2,0)] indicating a $\ket{T}$ ($\ket{S}$) spin state [Fig.~\ref{fig:3}(b)].

The read-out fidelity is limited by several factors. A histogram of the integrated sensing dot signal at time $\tau_\textsc{m}$ constructed from 10000 single-shot measurements with a random initial spin state shows two peaks, corresponding to the signal measured for each of the spin states [Fig.~\ref{fig:3}(d)]. Due to noise in the current traces, there is a small overlap between the two peaks that will lead to spin read-out errors. From a double Gaussian fit to the histograms, we extract an error contribution $e_\textsc{n} = 0.006$. When averaging 10000 complete read-out traces, the sensor signal shows an exponential decay, with a time constant $T_1$ [Fig.~\ref{fig:3}(c)]. The $T_1$ value varies with inter-dot detuning~\cite{PSB}, reaching up to $T_1 = 11.4$~$\mu$s. A relaxation event before $\tau_\textsc{m}$ leads to a measurement error so it is important to keep $\tau_\textsc{m}$ short. In order to achieve a sufficient signal-to-noise ratio, we low-pass filtered the signal with a 1~MHz cut-off, which in turn leads us to choose $\tau_\textsc{m} = 0.8$~$\mu$s. The $\ket{T}$ measurement error due to relaxation is then $e_\textsc{t1} = 1 - \exp(-\tau_\textsc{m} / T_1) = 0.07$. This is the dominant source of error in this system, with smaller error contributions from thermal excitation, limiting the average measurement fidelity to $F_\textsc{m} \approx 0.96$.

In summary we have implemented and operated a quantum dot plaquette with reliable control of electron filling and tunnel coupling, and for which we demonstrated single-shot spin measurement. This makes this system a promising analog quantum simulator of Fermi-Hubbard physics. The two-dimensional lattice configuration presents symmetries not accessible in the more common linear arrays, enabling the emulation of phenomena such as Nagaoka ferromagnetism~\cite{Nagaoka} and resonating valence bond states~\cite{ANDERSON}, which have been predicted for high-temperature superconductors. Moreover using the two-layer fabrication technique shown here, the 2$\times$2 geometry can be extended directly to a ladder of quantum dots (size 2$\times$N), which is the smallest system capable of showing pairing in under- or over-doped lattices~\cite{Dagotto618} and other interesting quantum phases~\cite{Bose}. Moreover, leveraging the fabrication experience of the semiconductor industry, quantum dot arrays might be scaled up to N$\times$N arrays, opening up a host of possibilities.
\\\\
Raw data and analysis files supporting the findings of this study are available from \url{https://doi.org/10.5281/zenodo.1219088}.

\begin{acknowledgments}
We acknowledge useful discussions with T. Hensgens, L. Janssen, T. Baart, T. Fujita, other members of the Vandersypen group, and J. Watson, V. Michal, as well as experimental support by C. van Diepen, P. Eendebak, R. Schouten, R. Vermeulen, R. van Ooijik, H. van der Does, M. Ammerlaan, J. Haanstra, S. Visser and R. Roeleveld. This work is supported by the Netherlands Organization for Scientific Research (FOM projectruimte and NWO Vici) and the Swiss National Science Foundation.
\end{acknowledgments}

\bibliography{citation}

\end{document}